\documentclass[12pt,a4paper]{article}

\usepackage{epsfig}
\usepackage{amsmath}

\def\nostrocostrutto#1\over#2{\mathrel{\mathop{\kern 0pt \rlap 
  {\raise.2ex\hbox{$#1$}}}
  \lower.9ex\hbox{\kern-.190em $#2$}}}
\def\gsim{\nostrocostrutto > \over \sim}   

\catcode`@=11
\newcount\@tempcntc
\def\@citex[#1]#2{\if@filesw\immediate\write\@auxout{\string\citation{#2}}\fi
  \@tempcnta\z@\@tempcntb\m@ne\def\@citea{}\@cite{\@for\@citeb:=#2\do
    {\@ifundefined
       {b@\@citeb}{\@citeo\@tempcntb\m@ne\@citea\def\@citea{,}{\bf ?}\@warning
       {Citation `\@citeb' on page \thepage \space undefined}}%
    {\setbox\z@\hbox{\global\@tempcntc0\csname b@\@citeb\endcsname\relax}%
     \ifnum\@tempcntc=\z@ \@citeo\@tempcntb\m@ne
       \@citea\def\@citea{,}\hbox{\csname b@\@citeb\endcsname}%
     \else
      \advance\@tempcntb\@ne
      \ifnum\@tempcntb=\@tempcntc
      \else\advance\@tempcntb\m@ne\@citeo
      \@tempcnta\@tempcntc\@tempcntb\@tempcntc\fi\fi}}\@citeo}{#1}}
\def\@citeo{\ifnum\@tempcnta>\@tempcntb\else\@citea\def\@citea{,}%
  \ifnum\@tempcnta=\@tempcntb\the\@tempcnta\else
   {\advance\@tempcnta\@ne\ifnum\@tempcnta=\@tempcntb \else \def\@citea{--}\fi
    \advance\@tempcnta\m@ne\the\@tempcnta\@citea\the\@tempcntb}\fi\fi}
\catcode`@=12

\begin{document}

\setcounter{page}{0}
\thispagestyle{empty}

\title{
Angular Distribution of Particle Flows and Perturbative QCD Predictions
\\  
\vspace{2cm}}

\author{Matthew A. Buican$^1$, 
\ Valery A. Khoze$^2$ 
\ and  \ Wolfgang Ochs$^1$ 
}

\date{
{\normalsize\it $^{1)}$Max-Planck-Institut 
f\"ur Physik, (Werner-Heisenberg-Institut) \\
F\"ohringer Ring 6, D-80805 M\"unchen, Germany\\
$^{2)}$Department of Physics and Institute for Particle Physics
Phenomenology\\
University of Durham, Durham, DH1 3LE, UK \\ \vspace{2cm}
}}

\maketitle

\thispagestyle{empty}

\begin{abstract}
We discuss the predictions of
perturbative QCD for angular flows of final state particles 
in two and three jet events
including their $cms$ energy and jet resolution ($y_{cut}$) 
dependence. The simple
analytical formulae for gluon bremsstrahlung from primary partons,
modified for gluon cascading,
reproduce the main features of the experimental data well.
For $y_{cut}$-selected events, the particle flow is derived from a
superposition of colour dipoles in much the same way that photon radiation 
is derived from electric dipoles.
\end{abstract}

\vspace{-20cm}~

\rightline{MPI-PhT/02-2002-78}
\rightline{IPPP/02/80; DCPT/02/160}

\newpage
\section{Introduction}
We study the angular distribution of particles between the hadronic
jets in the final state of a high energy collision. The
angular flow of such particles, most of them with low
momentum $<1$ GeV, is expected to depend characteristically on the 
colour connections of the primary partons \cite{bcs,adkt,dktm}
(for recent reviews, see, for example, Ref. \cite{ko}).
An early example of such a phenomenon concerns the final state
 $e^+ e^- \rightarrow q\overline{q}g$.
 The soft radiation is not distributed
symmetrically between the three jets,
but rather is depleted in the angular region 
between the $q$ and $\overline{q}$ directions. This effect was
first predicted within the string hadronization model
 \cite{ags}. Subsequently, it has been derived within 
perturbative QCD  \cite{adkt2} where 
the particle flow is derived directly from the soft gluon
bremsstrahlung, which is emitted coherently from all primary
partons but with different strength from gluon and quark emitters
according to the QCD colour factors.

The first observation of the effect was due to the 
JADE collaboration \cite{jade}
with many of the subsequent details having been studied by various experimental groups,
for example \cite{TPC,DELPHI,opal12}.
This \lq\lq string/drag effect''
 is by now well established and reproduced by the
popular Monte Carlo models. The analytic calculations have been verified
mainly for angles midway  between the jets. 
Until now, the full angular pattern predicted
by perturbative QCD had  not been compared systematically with data. At PEP energies
the TPC collaboration \cite{TPC} found that their data -- though in
qualitative agreement with QCD expectations -- showed some quantitative
differences with the asymptotic analytical results, which could have possibly
been
caused by the low purity 
of the quark/gluon identification of the jets.
At LEP there has been considerable progress in jet identification, in
particular by DELPHI \cite{DELPHI}; their result for the \lq\lq Mercedes 
configuration''
has been compared successfully \cite{ko} with the analytic 
calculation~\cite{adkt2,dkmt2}.
Comparison with the limited results by OPAL \cite{opal12}, averaged over a large range of jet
angles, has been moderately successful.

In this paper we compare the experimental results 
for the various symmetric and asymmetric angular configurations 
of $q\bar q g$ presented 
by DELPHI and OPAL for the full angular region,
along with the results for $q\bar q\gamma$ and two jet
events with theoretical predictions. Furthermore, we discuss the
energy and jet resolution dependence of the effect 
which has been ignored in the previous analyses.

\section{Angular pattern of particle flows}
\subsection{Perturbative results and dependence on the jet selection}
Let us begin by recalling the main ideas \cite{adkt2} in the 
perturbatively-based calculation
of particle flows in  $e^+ e^- \rightarrow
q\overline{q}g$ at high energies in comparison with  $e^+ e^- \rightarrow q\overline{q}
\gamma$.
We consider the configurations where  all the angles
$\Theta_{ij}$ between jets are large
($i=\{+-1\}\equiv \{q \overline q g\}$ or $i=\{+-\}\equiv \{q \overline q\}$).
Within
the perturbative picture, the angular distribution of soft inter-jet
hadrons is calculated from the distribution of soft gluons radiated
coherently
off the colour antenna formed by the primary emitters $(q, \overline{q}$ and
$g$) or $(q$ and $ \overline{q})$.  

The angular distribution of a secondary
soft gluon, $g_2$, 
is derived in lowest order perturbation theory from the corresponding
Feynman diagrams of single gluon emission off the primary partons.
For the $q\bar q \gamma$ process one finds  neglecting the
recoil
\begin{equation}
\frac{8 \pi d N_{q\overline{q}\gamma}}{d \Omega_{\vec{n_2}}E_2dE_2} =
\frac{4C_F \alpha_s(k_t)}{\pi}\;  
\frac{(p_+ p_-)}{(p_+k_2)\; (p_-k_2)} 
\label{born}
\end{equation}
with 4-momenta $p_+,p_-$ and $k_2$ and soft gluon energy 
and transverse momentum $E_2$ and $k_t$. 
 
A  modification is necessary to take into account the fact that the \lq\lq 
detected'' gluon belongs to a parton jet.
The results of the calculation for $q\overline q \gamma$ 
can be written as
\begin{equation}
\frac{8 \pi d N_{q\overline{q}\gamma}}{d \Omega_{\vec{n_2}}}=
\frac{1}{N_C} \; W_{+ -} (\vec{n}_2) \: N_g^\prime (Y) \; = \; 
\frac{2C_F}{N_C} \; (\widehat{+ -}) \:  N_g^\prime (Y),
\label{emit2}
\end{equation}
and for $q\overline q g$ one finds
\begin{equation}
\frac{8 \pi d N_{q\overline{q}g}}{d \Omega_{\vec{n}_2}} =
\frac{1}{N_C} \; W_{\pm 1} \: (\vec{n}_2) \: N_g^\prime (Y) \;
= \; \left [ ( \widehat{1 +}) + ( \widehat{1 -}) \: - \:
\frac{1}{N_C^2} \; ( \widehat{+ -}) \right ] \; N_g^\prime (Y).
\label{emit3}
\end{equation}
Here the angular distribution of soft 
gluon bremsstrahlung  from the
\lq\lq antenna''-dipole $(\widehat{i j})$ is obtained from (\ref{born})
and reads
\begin{equation}
(\widehat{i j}) \; = \; \frac{a_{ij}}{a_i a_j}, \;\;\;\; a_{ij}
\; = \; (1 \: - \: \vec{n}_i \vec{n}_j), \;\;\;\; a_i \; = \; (1
\: - \: \vec{n}_2  \vec{n}_i).
\label{4.2}
\end{equation}
This angular distribution is the same as that for photon bremsstrahlung in QED
-- as it occurs, for example, 
in $e^+ e^-$ pair creation.
Unlike the QED case, however, there are also the QCD-specific colour factors and the 
\lq\lq cascading factor''  $N_g^\prime (Y)$, which takes into account the
fact that $g_2$ is part of a jet.
It represents the derivative of particle multiplicity in a gluon jet
with respect to $Y=\ln(K_T/Q_0)$ at the appropriate maximal 
transverse momentum scale
$K_T$ and cut-off $Q_0$.
The parameter $Q_0$ is determined 
from a fit to the multiplicity 
data together with the QCD scale $\Lambda$ ($Q_0\gsim \Lambda$, see, for
example \cite{ko}).

The scale $K_T$ depends on the way the jets are selected. To see this, we write
down the perturbative expansion for the emission of 
gluon $g_2$  in (\ref{emit2}) 
from the primary parton $p$ as follows (see also \cite{dmo})
\begin{equation}
\begin{split}
\frac{8 \pi d N_{q\overline{q}\gamma}}{d \Omega_{\vec{n_2}}} 
& = \frac{1}{N_C} \; \int^{E_{max}}_{E_{min}}  \frac{dE_2}{E_2} 
        W_{+-}({\vec{n_2}}) \gamma_0^2(k^t_{2p})\\
&  \quad +\frac{1}{N_C} \; \int^{E_{max}}_{E_{min}}  \frac{dE_2}{E_2} 
        \int^{E_{max}}_{E_2}  \frac{dE_a}{E_a} 
             \int \frac{d\Omega_{ap}}{2\pi \Theta_{ap}^2} 
           \gamma_0^2(k^t_{ap}) W_{+-}({\vec{n_a}}) 
            \gamma_0^2(k^t_{2a}) + \ldots\\
&\approx  \frac{1}{N_C} \; 
        W_{+-}({\vec{n_2}})  \int^{E_{max}}_{E_{min}}  \frac{dE_a}{E_a}
        \gamma_0^2(k^t_{2p})
          [1+ \int^{E_a}_{E_{min}} \frac{dE_2}{E_2} \int 
          \frac{d\Omega_{a2}}{2\pi\Theta_{a2}^2} 
        \gamma_0^2(k^t_{a2}) + \ldots].
\label{expand}
\end{split}
\end{equation}
The first term corresponds to the Born result for direct gluon emission
in (\ref{born}), the second one to
the emission through the intermediate parton $a$.
We denote by 
 $k^t_{ab}$  and  $\Theta_{ab}$ 
the transverse momentum and angle respectively of parton $a$ 
with respect to parton $b$, in addition
$\Theta_{ap}\equiv \Theta_a$; the leading order multiplicity anomalous
dimension is given by $\gamma_0^2(k_t)=2N_C\alpha_s(k_t)/\pi$. 
Eq. (\ref{expand}) is written in DLA,
but it can also be generalized to MLLA after inserting the appropriate
splitting functions in all intermediate branchings.

In the second equation in (\ref{expand}) we have replaced the integral over the angle
$\Theta_{ap}$ by the integral over $\Theta_{a2}$, as the leading contribution
to the $a$ integral comes from the region of quasi-collinear emission
$\vec{n_a}\simeq \vec{n_2}$; then we also replace the corresponding angles
involving particle $a$ in $W_{+-}$. Furthermore, the order of integration
between $E_a$ and $E_2$ is interchanged ($E_2<E_a$).  Now, the angular integral in the second
term has to satisfy the ``angular ordering'' requirement, i.e. the
emission angle of the intermediate parton $a$ should respect 
\begin{equation}
\Theta_{2a}< \Theta_{ap};\quad \text{therefore, also}\quad 
                   \Theta_{ap}>\Theta_{2p}/2.
\label{angleorder}
\end{equation}
Then the angular integral can be performed with $\Theta_{a2}<\Theta_2$
in pole approximation.\footnote{The discussion
of boundaries in these integrals is the same as in the derivation of the
inclusive spectra, see for example, Ref. \cite{ow}.}

Next we discuss the limits of the energy integrations
in (\ref{expand}). The lower limit
$E_{min}$ is determined by the transverse momentum cut-off. For gluon
emission inside a well separated jet one requires $k_t>Q_0$, therefore
$E_{min}\approx Q_0/\Theta_2$ in the small angle approximation. In case of
 emission from a boosted $q\overline q$ ``dipole'' as in (\ref{born})
an azimuthal dependence of the effective cut-off $Q_0$ around the jet axis 
is generated. 
To estimate this effect we consider a generalization of the cut-off
restriction following from (\ref{born}) 
\begin{equation}
\tilde{k}_t^2=\frac{2(p_+k)(p_-k)}{p_+p_-}=
\frac{2E_2^2(1-\cos\Theta_{+2})(1-\cos\Theta_{-2})}{1-\cos \Theta_{+-}}
   \ > \ Q_0^2,
\label{ktlor}
\end{equation}
which approaches 
the limit $\tilde{k}_t=k_t$ for the $q\overline q$ back-to-back
 configuration. For our applications
the effective $k_t$ cut-off $Q_0$ will then change by up to a factor 2.
In the asymptotic expansion of multiplicity \cite{webber,adkt} $\ln N_g(Y)=c_1 \sqrt{Y}
+c_2\ln Y + c_3/\sqrt{Y}$  
such a rescaling $Q_0\to \beta Q_0$ would modify only the $c_3$ term,
therefore the leading (DLA) and next-to-leading (MLLA) results remain
unchanged. Numerically, for the present analysis we estimate the effect on 
$N'$ in (\ref{emit2}),(\ref{emit3}) from the variation of $Q_0$ 
to be at the 10-20\% level where we take into account the data in Fig. 4a
below. In our application of analytic high energy approximations 
(large $K_T/Q_0$) we continue 
therefore with the cut-off $k_t>Q_0$ in the following.

The upper limit $E_{max}$ of the energy integral depends on the jet
selection procedure and we consider two cases\\
{\it a) no momentum restriction in jet selection}\\
In the simplest case, we require 2 or 3 respectively 
energetic particles in the given angular directions which define the jet
directions. A more precise definition is possible using the 
energy-energy-multiplicity  or energy-energy-energy-multiplicity
correlations for 2 and 3 jets \cite{EMM,dktm}. For two jets one considers all 
pairs of particles (ij) weighted with their energies, and for a given relative 
angle $\Theta_{ij}$,
one studies the angular distribution of all soft particles in the event.
In the same way one proceeds for 3-jet events with triples of energy
weighted particles and two relative angles.

In these cases there is no specific restriction on the energy of the
triggered soft particles whose angular distribution is investigated. 
The maximum energy $E_{max}$ of these particles is 
then of order of the respective jet
energies $E_{jet}$, say $\sqrt{s}/2$ or $\sqrt{s}/3$ in 2 and 3 jet events at total cms 
energy $\sqrt{s}$ of the jet system, and the same applies to the intermediate 
partons $a$. 

{\it b) selection of jets for a given resolution parameter $y_{cut}$}\\
We consider here the ``Durham algorithm'' \cite{durham} which selects
the jets according to a predefined minimal relative 
transverse momentum, approximately
$K_T \geq \sqrt{y_{cut}s}$. In this case, the $y_{cut}$ selection 
restricts the transverse momenta of emitted
particles $g_2$ as well as the transverse momenta
of the intermediate partons $a$ in
(\ref{expand}) 
in the same way ($k^t_{ap}<K_T$), whereas
the longitudinal momenta of the latter partons $a$ are only 
limited by   ${\cal O}(\sqrt{s})$. 
However, an additional restriction comes from the angular ordering 
(\ref{angleorder}), which is
violated for intermediate partons $a$ with large energies
and, thus, small angles $\Theta_{ap}$. In our logarithmic
approximation we take the same upper limit for the $E_a$ and $E_2$ integrals.  
Similarly, the angular integral is limited by
$\Theta_{a2}<\Theta_{2}$ within the same accuracy. 
Then, the second equation in (\ref{expand}) can be resummed, resulting in the
multiplicity $N_g(K_T/Q_0)$  
\begin{equation}
\frac{8 \pi d N_{q\overline{q}\gamma}}{d \Omega_{\vec{n_2}}} 
\approx  \frac{1}{N_C} \; 
        W_{+-}({\vec{n_2}})  \int^{E_{max}}_{E_{min}}  \frac{dE_2}{E_2}
        \gamma_0^2(k^t_{2p}) N_g(k^t_{2p}/Q_0).
\label{multin}
\end{equation}
This integral can be expressed in terms of $N'(Y)$
using the evolution equation for multiplicity, which then leads to Eq.
(\ref{emit2}). We note that this replacement is also possible in
MLLA.\footnote{The MLLA evolution equation \cite{dkmt2} is usually written
in small angle approximation with $K_T=K\Theta$, the
continuation to larger angles is not unambiguous.} 

The upper
limit of integration $E_{max}$ is either given by the 
maximal energy $E_{jet}$ of the jet
or depends on the jet resolution parameter $y_{cut}$ 
\begin{eqnarray}
  \text{case a (no restriction)}\; & Y=  \ln (E_{jet} \Theta_m/Q_0)
    \label{Ymno}\\
  \text{case b ($y_{cut}$ restriction}) &  
  Y= \begin{cases} \ln (E_{jet} \Theta_m/Q_0) &
                                 \Theta_m<\sqrt{y_{cut}s}/E_{jet}
      \label{Ym} \\
                 \ln (\sqrt{y_{cut}s}/Q_0) & \text{otherwise}
 \end{cases}\\
\nonumber 
\end{eqnarray}
Here $\Theta_m$ is the angle between the soft gluon $g_2$ and the jet
closest in angle. 
As the multiplicity $N_g$ depends on the  maximum transverse
momentum scale, the Durham $K_T$\ -\ algorithm naturally 
provides simple results. Note that in case of $y_{cut}$ selection (case b),
the $N'(Y)$ factor becomes independent on the emission angle $\Theta_m$
away from the jet directions, and, therefore, the soft gluon angular
distribution is given simply by the Born-term factors $W(\vec{n_2})$
in (\ref{emit2},\ref{emit3}).

\subsection{Projection onto the event plane}
\begin{figure}[t!]
\begin{center}
\mbox{\epsfig{file=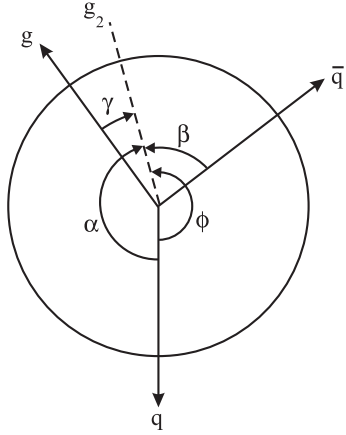,width=4.5cm}}
\end{center}
\vspace{-0.3cm}
\caption{
Definition of azimuthal angles in the event plane of three jet event.
}
\label{anglefig}
\end{figure}

We are now interested in the projection of the soft gluon $g_2$
 angular distribution onto the event plane defined by the momentum vectors
of the $q\bar q g (\gamma)$ systems. The azimuthal angle $\phi$ in the plane  
is defined to be zero in the direction of $q$, becoming 
positive in the direction
of $\bar q$ for definiteness 
(in our applications $q$ and $\bar q$ are not
distinguishable), see
Fig.~\ref{anglefig}. The projection 
of the particle density onto the event plane
is obtained in the case of $q\overline{q}\gamma$ from the integral over
$\cos \Theta_2$, where $\Theta_2$ is the polar angle with respect to 
the normal to the event plane,
\begin{eqnarray}
W_{+ -} (\phi) & = & 2C_F \; \int \; \frac{d \cos \Theta_2}{2}
\; 
(\widehat{+ -}) \; = \; 2C_F \: a_{+ -} \: V (\alpha, \beta),
\label{projqqga} \\ 
V (\alpha, \beta) & = & \frac{2}{\cos \alpha - \cos \beta} \:
\left ( 
\frac{\pi - \alpha}{\sin \alpha} \: - \: \frac{\pi - \beta}
{\sin \beta} \right),
\nonumber
\end{eqnarray}
the angles $\alpha,\beta$ are given below in terms of $\phi$.

Replacing  $\gamma$ by the hard gluon $g_1$ one obtains an additional particle flow in
the $g_1$ direction, and from the integration of Eq.\ (\ref{emit3}) over
cos$\Theta_2$ one derives
\begin{equation}
W_{\pm 1} (\phi) = N_C \: \left [ a_{+1} V (\alpha, \gamma) +
a_{-1} V (\beta,
\gamma) - \frac{1}{N_C^2} \: a_{+ -} V (\alpha, \beta) \right ].
\label{projqqg}
\end{equation}
For an arbitrary choice of $\phi$ 
the angles $\alpha,\beta,\gamma$ in the  projection formulae
(\ref{projqqga}),(\ref{projqqg}) have to be defined as minimal angles
 between the directions of 
the soft gluon ($g_2$) and the directions of $q,\bar q$ and $g$ respectively,
with all angles within the interval $[0,\pi]$. 
We find the following definition of the angles $\alpha,\beta,\gamma$
to satisfy this requirement and to apply in all angular sectors with
$\phi$ running from $0$ to $2\pi$  (see also Fig.~\ref{anglefig})
\begin{eqnarray}
\alpha&=&\min(\phi,2\pi-\phi);\quad   
\beta =\min(|\Theta_{+-}-\phi|,2\pi -\phi+\Theta_{+-});\\
\gamma&=&\min(\phi+\Theta_{1+}, |2\pi-\Theta_{1+}-\phi|).
\label{angles} 
\end{eqnarray}

Eq. (\ref{projqqg}) corresponds to an incoherent superposition of two
Lorentz-boosted dipoles, 
one between the gluon and the quark and one between the gluon and
 the anti-quark,
modified by the negative correction of ${\cal O} (1/N_c^2)$. This expression
shows  a depletion opposite to the gluon direction.

\subsection{Comparison with experiment}
\begin{figure}[t!]
\begin{center}
\mbox{\epsfig{file=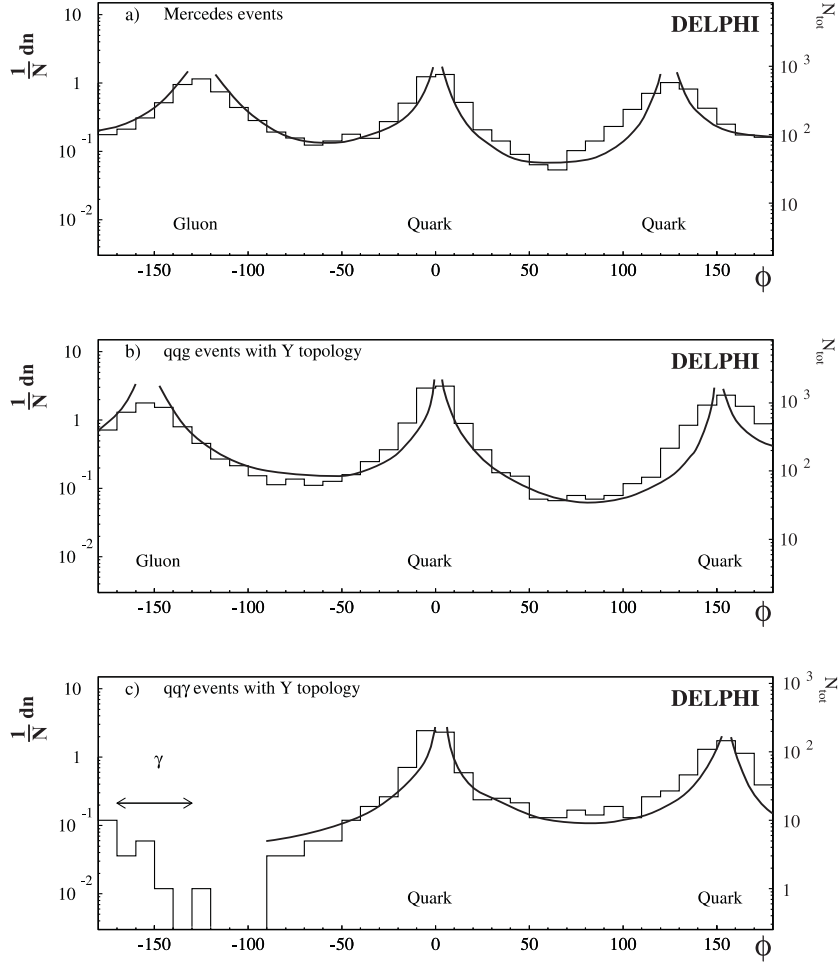,width=11cm}}
\end{center}
\vspace{-0.3cm}
\caption{
Charged particle flow within various multi-jet  configurations as measured by 
DELPHI  \protect\cite{DELPHI}
(a) \lq\lq Mercedes'' $q\bar q g$ events, (b) \lq\lq Y-symmetric'' $q\bar q g$ events 
and c) $q\bar q
\gamma$ events, in comparison with analytical QCD predictions; the curves
correspond to the lowest order QCD soft bremsstrahlung formulae.
The relative angles between the jets are taken in a) as 
 $\Theta_{1+}=125^\circ$, $\Theta_{+-}=122.5^\circ$ and  
$\Theta_{1-}=112.5^\circ$; in b) and c) as
$\Theta_{+-}=\Theta_{1+}=150^\circ$.
}
\label{delphiplot}
\end{figure}

Let us now compare these analytical
formulae with experimental data. 
The DELPHI collaboration \cite{DELPHI} 
has presented results on three jet events with two
identified $b$-quark jets allowing for a high gluon jet purity ($\sim 94\%$).
Symmetric angular configurations have been selected -- either \lq\lq Mercedes''
or \lq\lq Y-symmetric'' events with angles confined to an angular region of
$\pm15^\circ$. Furthermore, the data on the 
final state $q\bar q \gamma$ in the
Y-configuration have been presented. 
These well defined event classes 
are compared in Fig. \ref{delphiplot} with the above analytical formulae. 

The angle $\phi=0$ is in the direction of the most energetic quark jet.
For Mercedes events, we chose the other jet angles at
the observed peak positions -- slightly
different ($<8^\circ$) from the symmetric values.
The events had been selected with fixed  $y_{cut}\simeq 0.01$ 
according to the Durham algorithm. As a result,
we can neglect the variation of the cascading factor $N_g^\prime$ in 
(\ref{emit2}) and (\ref{emit3}) (for angles $\gsim 20^\circ$ away from the jet
directions). The shape of the angular distribution is then predicted without 
a free parameter; in  Fig. \ref{delphiplot} only one overall normalization
factor has been fitted in the description of the three distributions.

The general characteristics of the distributions are well reproduced, in
particular the different heights of the inter-jet valleys. In addition, the 
larger width of the gluon jet in comparison with the central quark jet width 
becomes
visible. The quark-jets in a) and b) are a bit broader than 
perturbatively predicted, 
as can be expected from the b-quark jet selection; 
the quark-jets at the 
larger angle $\phi$  in a) and b) are even broader because of
the angular fluctuation in the selection of the jets. This successfull
description nicely demonstrates the similarity of particle flow in \lq\lq
strong interaction'' processes to photon flow in electromagnetic
interactions -- both are caused by the 
underlying gauge boson bremsstrahlung in QED and QCD. 
In c) the formulae are equivalent,
in a) and b) they are modified by the 
different colour charges for quarks
and gluons -- a degree of freedom not
available in QED. An interesting QCD interference effect is the negative
$1/N_C^2$ term in (\ref{emit3}). Unfortunately, according to our analysis, 
the presented data are not
sufficient for a reliable observation of this term.

A comparison of $q\bar q g$ and  $q\bar q \gamma$ distributions between 
the quark and anti-quark jets has been presented by OPAL \cite{opal12}.
In the analysis of  $q\bar q g$ events, 
the jets are ordered according to their energies. 
The jet with the lowest energy in the $q\bar q g$ system is taken as
the gluon jet and two intervals of this energy $E_3$ are selected
around  $E_3=10$ and  $E_3=20$~GeV, furthermore, 
 $\Theta_{+-}\simeq 165^\circ$ from which we can estimate  $\Theta_{1-}$ and
 $\Theta_{1+}$.
In the calculation we take
into account the estimated purities of gluon jets, $p_g$, by superimposing
the distribution (\ref{emit2}) with the one 
obtained by swapping the second and third
jet, i.e. by swapping $\Theta_{+-}$ and  $\Theta_{1+}$ and changing $\phi$
to $2\pi - \phi$. The original distribution gets the weight $p_g=0.92 $ 
for the sample with $E_3=10$ and  $p_g=0.74 $ for
$E_3=20$ GeV, according to the reported purities.

\begin{figure}[t!]
\begin{center}
\mbox{\epsfig{file=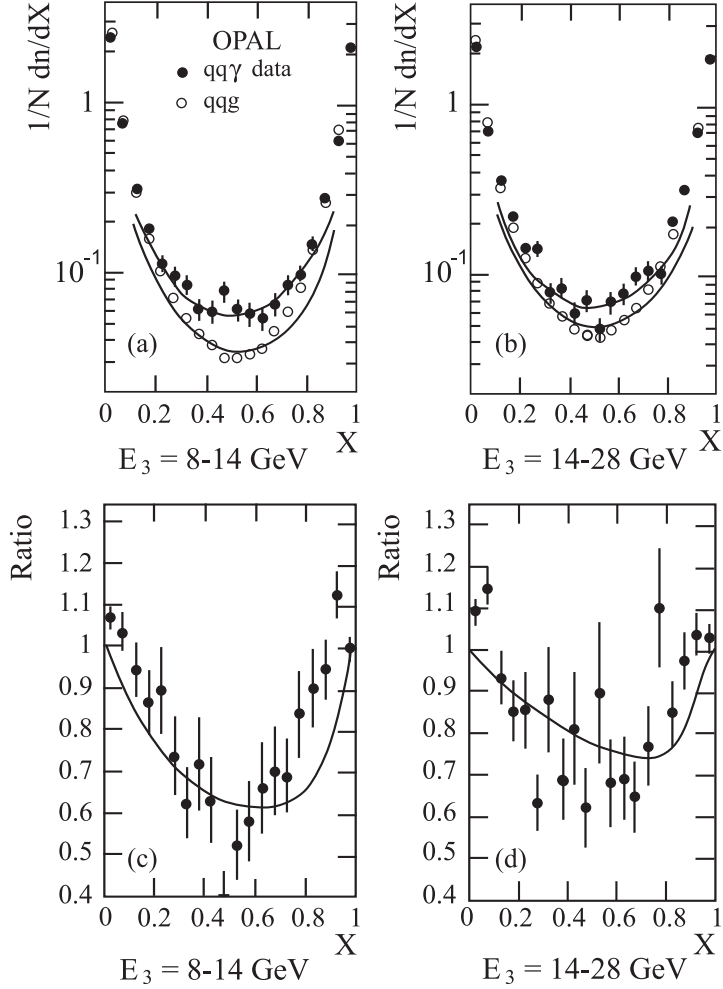,width=9.5cm}}
\end{center}
\vspace{-0.5cm}
\caption{
Charged particle flow in between the two quark jets of the $q\bar q g$ and
$q\bar q \gamma$ final states, as measured by OPAL  \protect\cite{opal12},
for two samples of energies $E_3$ of the lowest
momentum jet together with the corresponding ratios. The curves represent
the lowest order QCD bremsstrahlung formulae for two intervals of
 the lowest energy  $E_3$ of the jets; 
the
overall normalization has been adjusted; in addition, 
the curves in b) are increased by 15\% (see text).  
 In our calculations we use for the first
interval $E_3=10$~GeV, $\Theta_{+-}= 165^\circ$, $\Theta_{1-}= 67^\circ$
and  $\Theta_{1+}= 128^\circ$, and for the second one 
$E_3=20$~GeV, $\Theta_{+-}= 165^\circ$, $\Theta_{1-}= 35^\circ$
and  $\Theta_{1+}= 160^\circ$. 
}
\label{figopal}
\end{figure}

In Fig. \ref{figopal} we show the distribution in the rescaled azimuthal angle
$X=\phi/\Theta_{+-}$  for the two processes 
as well as for their ratios in the two intervals of $E_3$.     
The different heights of both distributions  near $X\simeq 0.5$ is rather well
reproduced, as is the variation with
energy $E_3$ and jet angle $\Theta_{1-}$.
In particular, there is a sizeable asymmetry in the distribution which would be
even more pronounced if the gluon jets
were identified with higher purity.
Since the data in both intervals still average a sizable range of different
angular configurations, some deviations from the predictions are to be expected. 
We
therefore allowed an increase of relative normalization by 15\% of the
curves in Fig  \ref{figopal} b) over  Fig  \ref{figopal} a) to improve the
description.
The predictions are seen to 
fall below the data for $dn/dX$ near the jet directions,
which may be due to the averaging over the jets with different
energies and angular widths.
 On the other hand, the difference in the 
splitting of the two distributions, in their asymmetry,
 and in their absolute height for the two angular configurations in a) and b)
are well reproduced.
 The effects related to configuration asymmetry 
have been studied here for the first time; these effects
could be investigated more precisely in the future if smaller intervals in 
the relative jet angles were chosen.

\section{Energy dependence of particle angular flow}
The energy ($\sqrt{s}$) and $y_{cut}$ dependences  
are determined by the cascading factor $N'(Y)$ according to
(\ref{Ymno}),(\ref{Ym}).
Angular distributions for particles produced in the hemispheres of 
quark and gluon jets 
 without  $y_{cut}$ restriction are available 
at different energies from $e^+e^-\to q\bar q$ and at LEP-1 energy
from  $e^+e^-\to b\bar b g$.
Instead of the distribution in angle $\Theta$ we may also study the
distribution in pseudo-rapidity $y=-\ln\tan (\Theta/2)$.
For quark and gluon jets ($A=q,g)$ we have
\begin{equation}
\frac{dn_A}{dy}(E_{jet},y) = N'_A(Y),\quad
Y=\ln\left(\frac{E_{jet}\Theta}{Q_0}\right). 
\label{angleNp}
\end{equation}
This relation is not limited to the leading order result $N_q/N_g=C_F/C_A$
as in Eq. (\ref{emit2}) but allows for higher order logarithmic effects.
Relation (\ref{angleNp}) follows from the MLLA result that the
multiplicity depends only on the maximum $K_T$ in the jet, i.e. on the
product $E_{jet}\Theta$ in the small angle approximation.\footnote{
Note that the remarkable MLLA prediction \cite{adkt,dktm} of such
$E_{jet}\Theta$ scaling behaviour has been recently well confirmed
experimentally by the CDF collaboration \cite{cdf} in the measurements of
inclusive charged particle production in restricted cones around the jet
direction.} 
 Then, the variation of
multiplicity with $y\simeq -\ln (\Theta/2)$ 
is the same as the one with $\ln E_{jet} \simeq
 y_{max}=\ln(2E_{jet}/m)$ or, equivalently, the particle density depends
on rapidity and energy only through the scaling variable 
\begin{equation}
x=y-\ln(\sqrt{s}/\mu)\ = \ -\ln(E_{jet}\Theta/\mu)
\label{scaly}
\end{equation}
where we set $\mu=1$ GeV and $\sqrt{s}=2E_{jet}$. 
Eq. (\ref{angleNp}) predicts both the energy
dependence at fixed angle (rapidity)
and the angular dependence at fixed
energy. Recall that within the MLLA approach \cite{adkt,dkmt2} the multiplicity
$N$ depends only on the phenomenological parameter $Q_0\sim\Lambda$ and a
normalization factor, then the angular distribution is uniquely
predicted by (\ref{angleNp}). 

\begin{figure}[t!]
\begin{center}
\mbox{\epsfig{file=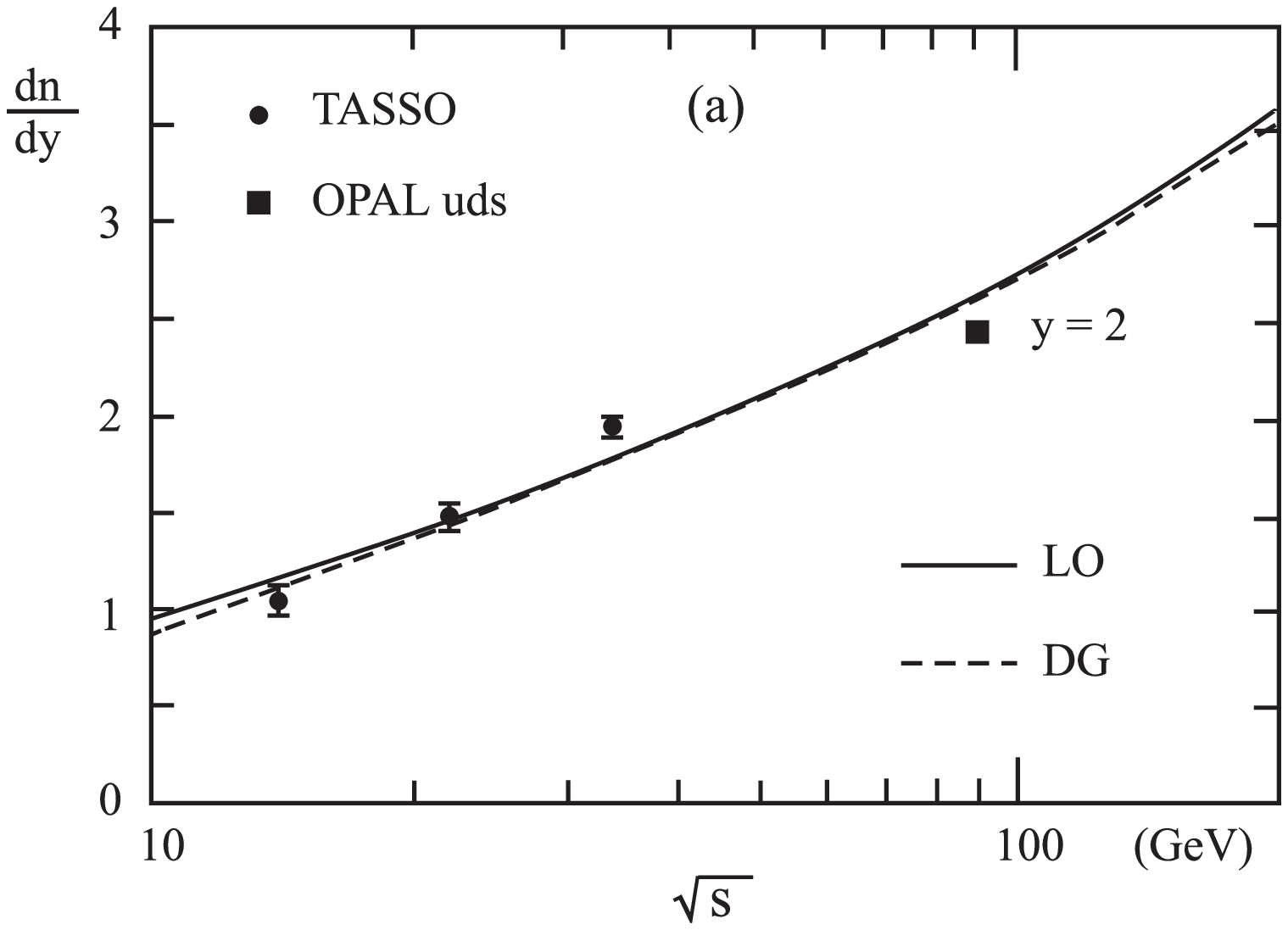,width=8.5cm}}\\
\mbox{\epsfig{file=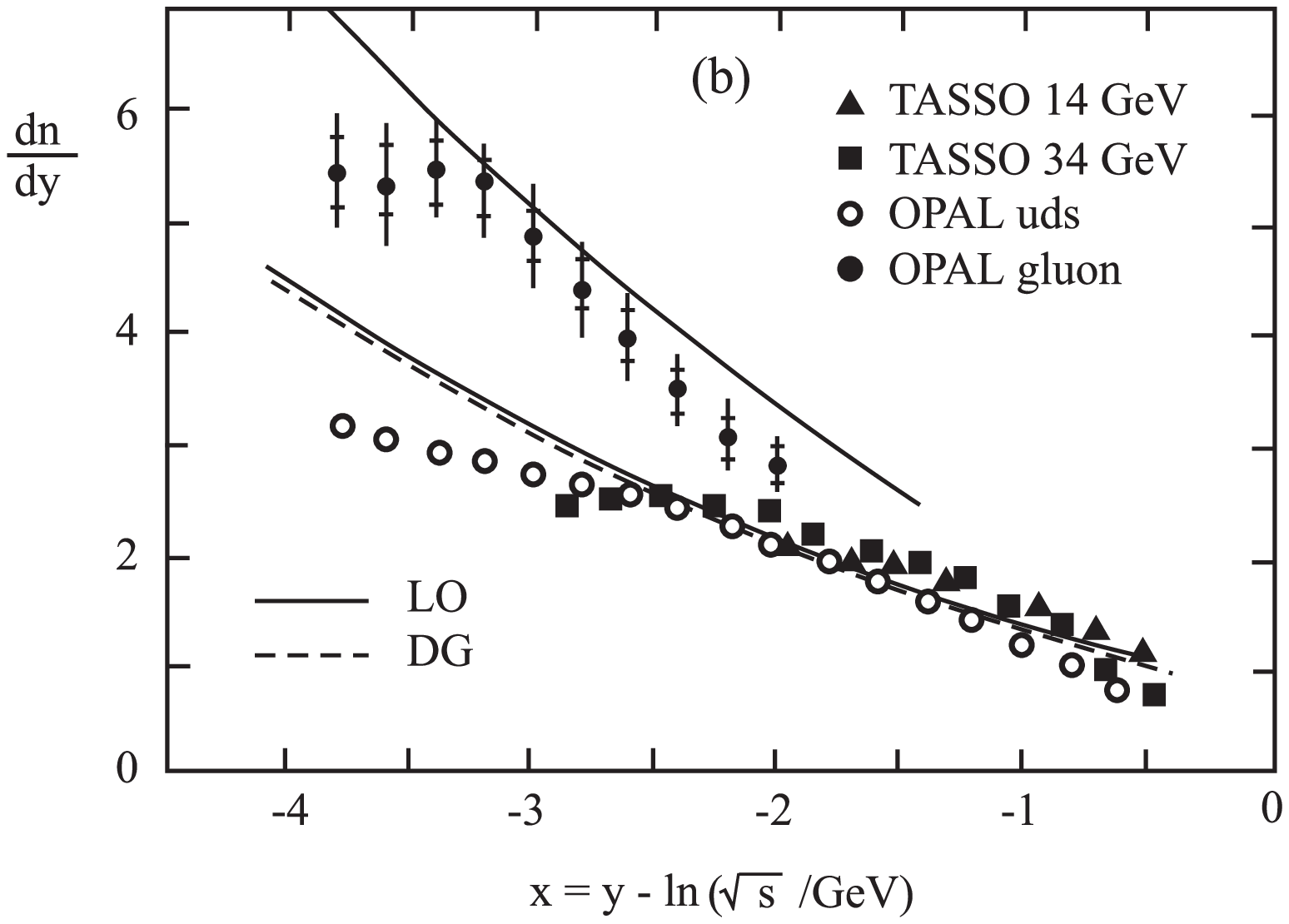,width=8.5cm}}
\end{center}
\vspace{-0.6cm}
\caption{
Distribution in rapidity $y$ (a) at fixed rapidity $y=2$ versus energy 
$\sqrt{s}$  and
(b) as a function of the shifted rapidity $x=y-\ln(\sqrt{s}/{\rm GeV})$
at different energies $\sqrt{s}$ (assuming pion mass for all charged
particles),
in comparison with solutions of the MLLA evolution equations
for $N'$ by Dremin and Gary (DG) 
\protect\cite{dg} and by Lupia and Ochs (LO) \protect\cite{lo}; data by
TASSO \protect\cite{tasso} and OPAL \protect\cite{opal}.
}
\label{figrap}
\end{figure} 

These predictions have been tested by comparing the QCD
results for $N'_A$ with data from TASSO and OPAL (here from $uds$-quark jets
only to avoid the heavy-quark contribution at the $Z$ mass) 
in Fig. \ref{figrap}. The curves in Fig. \ref{figrap} represent
the high energy QCD  3NLO
prediction (DG \cite{dg})\footnote{we selected the fit with parameters 
$n_f$=3, $\Lambda$=0.67 GeV and K=0.288 for a gluon jet}
and the numerical solution (LO \cite{lo}) of the full MLLA evolution equations
for quark and gluon jets \cite{dkmt2} . In both calculations the parameters
have been determined from a fit to $e^+e^-$ data, and the two results agree 
for quark jets within $\sim 3\%$ in the energy region considered
($10<\sqrt{s}<200$ GeV). The data in  Fig. \ref{figrap}a show the predicted
increase with energy. In  Fig. \ref{figrap}b we show the comparison for the
rapidity distributions at different energies using the scaling variable
$x$ in (\ref{scaly}). At each energy, the lower bound of the rapidity interval 
is $y = 0.8$. Above this value the scaling property is approximately satisfied,
and, again, a good description by the theoretical calculation of $N'(y)$ is
obtained in absolute terms. However, some discrepency at low $y$ is seen for the OPAL
data.\footnote{OPAL defines rapidity with respect to the sphericity
axis, while TASSO defines rapidity with respect to the thrust axis; 
without the $uds$ selection and using the thrust axis to define rapidity
ALEPH \cite{alephrev} data would lie above the curves.}
Approximate agreement is also obtained with the distribution in
gluon jets from OPAL \cite{opal} in the given rapidity range.
For larger rapidities, the distibution would fall below the distribution for
quarks as one may expect from energy conservation and recoil effects
 given the increased
central production of particles. The curve shown
corresponds to the common solution of the evolution equations 
for quark and gluon jets (LO \cite{lo}); 
the solution DG \cite{dg} with the quark jet parameters would 
result in a
prediction larger by 20\%; with readjusted parameters a good
description is obtained again. 

We conclude that the energy
dependence of the angular distribution, given by $N'(Y)$ in absolute terms,
 is reasonably well reproduced. A further improvement would require the
treatment of
(i) large angle and recoil corrections, (ii) dependence on jet axis
definition (iii) heavy quark
contribution and hadron mass effects.

The energy dependent effects discussed here are also expected
in multi-jet events as in Eqs. (\ref{emit2}, \ref{emit3}). 
Different analysis methods and
configurations selected by different groups do not allow a meaningful
comparison yet.
Another interesting test will be the dependence 
of angular distributions on the jet resolution parameter
$y_{cut}$ in (\ref{Ym}).

\section{Conclusion}
The angular distribution of particles in two and three jet events follows
closely the expectations based on the lowest order bremsstrahlung formulae of
QCD. These formulae can be represented as  sums over dipole radiation
terms (\ref{4.2}), the same as in QED, modified by 
the proper colour factors for quarks and
gluons. In addition, there is an effect of parton branching which modifies these
formulae by the ``cascading factor''  $N'(Y)$, the logarithmic
derivative of the multiplicity. For fixed resolution
parameter $y_{cut}$, 
this factor does not modify the basic angular
distribution in a wide range. On the other hand, it implies an additional  dependence 
on energy and/or jet resolution. The predicted dependence on $N'$ 
has been clearly established for two-jet events. These results demonstrate
that soft multi-hadron production follows the simple expectations of 
perturbative QCD bremsstrahlung and confirm the close connection between the
colour flows and observable flows of hadrons.

\section*{Acknowledgements}
We thank Bill Gary for useful discussions. VAK thanks the Leverhulme Trust
for a Fellowship and the theory group of the Max-Planck-Institute, Munich
for their hospitality.

\end{document}